\documentclass[12pt,titlepage]{article}
\usepackage[dvips]{epsfig}
\usepackage{cite}

\def\vek#1{{\bf #1}}
\def\veg#1{\mbox{\protect\boldmath $#1$}}
\def\LQCD{\Lambda_{\mathrm{QCD}}}
\def\gsim{\mathop{\raisebox{-.4ex}{\rlap{$\sim$}} \raisebox{.4ex}{$>$}}}
\def\lsim{\mathop{\raisebox{-.4ex}{\rlap{$\sim$}} \raisebox{.4ex}{$<$}}}

\begin{document}

\title{\vskip-2.0in{\normalsize\hfill FERMILAB-PUB-99/001-T}\vfill
Lattice QCD calculation of $\bar{B}\to Dl\bar{\nu}$ decay
form factors at zero recoil}

\author{
  Shoji~Hashimoto,$^{1,2}$
  Aida~X.~El-Khadra,$^3$ Andreas~S.~Kronfeld,$^1$\\
  Paul~B.~Mackenzie,$^1$ Sin\'{e}ad~M.~Ryan,$^1$ and
  James~N.~Simone$^1$\\[5mm]
  $^1$
  {\small\it Fermi National Accelerator Laboratory, 
  Batavia, IL 60510, USA}\\
  $^2$
  {\small\it 
  High Energy Accelerator Research Organization (KEK),
  Tsukuba 305-0801, Japan}\\
  $^3$
  {\small\it Physics Department, University of Illinois,
    Urbana, IL 61801, USA}
  }

\date{14 June 1999 \\ \small (Revised 27 October 1999)\vfill}
\maketitle

\begin{abstract}
A lattice QCD calculation of the $\bar{B}\to Dl\bar{\nu}$ decay form
factors is presented.
We obtain the value of the form factor $h_+(w)$ at the zero-recoil
limit $w=1$ with high precision by considering a ratio of correlation
functions in which the bulk of the uncertainties cancels.
The other form factor $h_-(w)$ is calculated, for small
recoil momenta, from a similar ratio.
In both cases, the heavy quark mass dependence is observed through 
direct calculations with several combinations of initial and final 
heavy quark masses.
Our results are $h_+(1) = 1.007(6)(2)(3)$ and
$h_-(1)=-0.107(28)(04)(^{10}_{30})$.
For both the first error is statistical, the second stems from the 
uncertainty in adjusting the heavy quark masses, and the last from 
omitted radiative corrections.
Combining these results, we obtain a precise determination of the 
physical combination $\mathcal{F}_{B\to D}(1)=1.058(^{20}_{17})$, 
where the mentioned systematic errors are added in quadrature.
The dependence on lattice spacing and the effect of quenching are not 
yet included, but with our method they should be a fraction 
of~$\mathcal{F}_{B\to D}-1$.
\end{abstract}

\newpage

\section{Introduction}
\label{sec:Introduction}

The precise determination of the Cabibbo-Kobayashi-Maskawa (CKM)
matrix element~$V_{cb}$ is a crucial step for $B$ physics to pursue
phenomena beyond the Standard Model.
In particular, the precision achieved in determining the apex of
the unitarity triangle may be limited by $|V_{cb}|$,
even with future high-statistics experiments.
The current determination of $|V_{cb}|$~\cite{Babar_98} is made
through inclusive~\cite{Ball_Beneke_Braun_95,Bigi_Shifman_Uraltsev_97}
and exclusive~\cite{Neubert_94,Shifman_Uraltsev_Vainshtein_95}
$B$~decays.

The heavy quark expansion offers a method to evaluate the
hadronic transition amplitude in a systematic way.
In particular, at the kinematic end point the exclusive $\bar{B}\to D^*$
matrix element is normalized in the infinite heavy quark mass limit,
and the correction of order $1/m_Q$ vanishes as a consequence
of Luke's theorem~\cite{Luke_90}.
It is thus possible to achieve an accuracy on~$|V_{cb}|$ of a few
percent.
Calculations of the $1/m_Q^2$ (and higher order) deviations from
the heavy quark limit have previously been attempted with the
non-relativistic quark model and with QCD sum rules.

Lattice QCD has the potential to calculate 
exclusive transition matrix elements from first
principles.
The shapes of the $\bar{B}\to D^{(*)}l\bar{\nu}$ decay
form factors have already been calculated successfully with
propagating~\cite{Bernard_Shen_Soni_94,UKQCD_95,Bhattacharya_Gupta_96},
static~\cite{Mandula_Ogilvie_94,Draper_McNeile_96,MILC_98,%
Christensen_Draper_McNeile_98},
and non-relativistic~\cite{Hashimoto_Matsufuru_96} heavy quarks.
On the other hand, a precise determination of the absolute normalization
of the form factors has not been achieved.
This paper fills that gap for the decay $\bar{B}\to Dl\bar{\nu}$.

Previous lattice calculations were unable to obtain the normalization of
the form factors for various reasons.
First, the statistical precision of the three point function
$\langle D V_{\mu} B^{\dagger}\rangle$, which is calculated by
Monte Carlo integration, has not been enough.
Second, perturbative matching between the lattice and the continuum
currents has been a large source of uncertainty.
Since the local vector current defined on the lattice is not a
conserved current at finite lattice spacing~$a$, the matching factor
is not normalized even in the limit of degenerate quarks.
Although one-loop perturbation theory works significantly better with
tadpole improvement~\cite{Lepage_Mackenzie_93}, the two-loop
contribution remains significant ($\alpha_s^2\sim$ 5~\%).
Last, the systematic error associated with the large heavy quark mass
must be understood.
Previous work with Wilson
quarks~\cite{Bernard_Shen_Soni_94,UKQCD_95,Bhattacharya_Gupta_96}, for
which the discretization error was as large as $O(am_Q)$, could not
address the $1/m_Q$ dependence in a systematic way when $m_Q\gsim 1/a$.

In this paper we present a lattice QCD calculation of the
$\bar{B}\to Dl\bar{\nu}$ decay form factor.
For the heavy quark we use an improved
action~\cite{Sheikholeslami_Wohlert_85} for Wilson fermions,
reinterpreted in a way mindful of heavy-quark
symmetry~\cite{El-Khadra_Kronfeld_Mackenzie_97}.
Discretization errors proportional to powers of~$am_Q$ do not exist in
this approach.
Instead, discretization errors proportional to powers of $a\LQCD$
remain, although they are intertwined with the~$1/m_Q$ expansion.
The first extensive application of this approach to heavy-light
systems was the calculation~\cite{Fermilab_97,JLQCD_97} of the
heavy-light decay constants, such as $f_B$ and $f_D$.
There the lattice spacing dependence was studied from direct
calculations at several lattice spacings, and a very small $a$
dependence was observed.
The third difficulty mentioned above is, thus, no longer a problem.

To obtain better precision on the semi-leptonic form factors, we
introduce ratios of three-point correlation functions.
The bulk of statistical fluctuations from the Monte Carlo integration
cancels between numerator and denominator.
Furthermore, the ratios are, by construction, identically one in both
the degenerate-mass limit and the heavy-quark-symmetry limit.
Consequently, statistical and all systematic errors, as well as the
signal, are proportional to the deviation from one.
The first and second difficulties given above are, thus, also
essentially cured.

The ratio of correlation functions for the calculation of $h_+(1)$
corresponds to the ratio of matrix elements,
\begin{equation}
  \label{eq:ratio_1}
  \frac{\langle      D |\bar{c}\gamma_0 b|\bar{B}\rangle
        \langle \bar{B}|\bar{b}\gamma_0 c|     D \rangle}
       {\langle      D |\bar{c}\gamma_0 c|     D \rangle
        \langle \bar{B}|\bar{b}\gamma_0 b|\bar{B}\rangle}
  = |h_+(1)|^2,
\end{equation}
in which all external states are at rest.
The denominator may be considered as a normalization condition
of the heavy-to-heavy vector current, since the vector current
$\bar{q}\gamma_{\mu}q$ with degenerate quark masses is conserved in
the continuum limit, and its matrix element is, therefore, normalized.
As a result the perturbative matching between the lattice and continuum
currents gives only a small correction to~$|h_+(1)|$.

For the calculation of $h_-(w)$ we define another ratio, corresponding
to matrix elements
\begin{equation}
  \label{eq:ratio_2}
  \frac{\langle D|\bar{c}\gamma_i b|\bar{B}\rangle}
       {\langle D|\bar{c}\gamma_0 b|\bar{B}\rangle}
  \frac{\langle D|\bar{c}\gamma_0 c|D\rangle}
       {\langle D|\bar{c}\gamma_i c|D\rangle}
  = 1 - \frac{h_-(w)}{h_+(w)},
\end{equation}
where equality holds when the final-state $D$ meson has small spatial
momentum.
By construction, the ratio produces a value of $h_-$ that vanishes
when the $b$ quark has the same mass as the $c$ quark, as required
by current conservation.

This method does not work as it stands for the
$\bar{B}\to D^*l\bar{\nu}$ decay form factors.
The axial vector current mediates this decay, and it is
neither conserved nor normalized.
We will deal separately with this case in another paper.

This paper is organized as follows.
Section~\ref{sec:form_factors} contains a general discussion of
form factors for the exclusive decay $\bar{B}\to Dl\bar{\nu}$.
Sections~\ref{sec:HQET_and_1/m_Q_expansion} and~\ref{sec:Lattice_and_HQ}
discuss heavy quark effective theory and the $1/m_Q$ expansion in
the continuum and with the lattice action used here.
Section~\ref{sec:Lattice_details} contains details of the numerical
calculations.
Sections~\ref{sec:Calculation_of_h_+}--%
\ref{sec:Heavy_quark_mass_dependence_of_h-} present our results.
Sections~\ref{sec:Calculation_of_h_+} and
\ref{sec:Heavy_quark_mass_dependence_of_h+}
discuss the form factor $h_+$ and its mass dependence.
Sections~\ref{sec:Calculation_of_h_-} and
\ref{sec:Heavy_quark_mass_dependence_of_h-} do likewise for $h_-$.
We compare the results from the fits of the mass
dependence to corresponding results from QCD sum rules in
Sec.~\ref{sec:Comparison_with_the_QCD_sum_rules}.
The values of $h_+(1)$ and $h_-(1)$ at the physical quark masses
are combined in Sec.~\ref{sec:Result_for_F} into a result for the
form factor $\mathcal{F}_{B\to D}(1)$, which with experimental data
determines~$|V_{cb}|$.
We give our conclusions in Sec.~\ref{sec:Conclusions}.

\section{$\bar{B}\to Dl\bar{\nu}$ form factors}
\label{sec:form_factors}

The decay amplitude for $\bar{B}\to Dl\bar{\nu}$ is parametrized with
two form factors $h_+(w)$ and $h_-(w)$ as
\begin{equation}
  \label{eq:definition_of_the_form_factors}
  \langle D(\vek{p}')| \mathcal{V}_{\mu} |
  \bar{B}(\vek{p})\rangle =
  \sqrt{m_B m_D}\left[h_+^{B\to D}(w) (v+v')_{\mu}
               + h_-^{B\to D}(w) (v-v')_{\mu}\right],
\end{equation}
where $v$ and $v'$ are the velocities of the $B$ and $D$ mesons,
respectively, and $w=v\cdot v'$.
The square of the momentum transferred to the leptons is then
$q^2  =  m_B^2 + m_D^2 - 2 m_B m_D w$.
We denote by the symbol~$\mathcal{V}_\mu$ the physical vector
current, to distinguish it from currents in heavy quark
effective theory (HQET) and in lattice QCD.

The differential decay rate reads
\begin{equation}
  \label{eq:differential_decay_rate}
  \frac{d\Gamma (\bar{B}\to Dl\bar{\nu})}{dw} =
  \frac{G_F^2}{48\pi^3}
  (m_B+m_D)^2 m_D^3 (w^2-1)^{3/2}
  |V_{cb}|^2 |\mathcal{F}_{B\to D}(w)|^2,
\end{equation}
with
\begin{equation}
    \label{eq:form_factor_relation}
    \mathcal{F}_{B\to D}(w) =
        h_+^{B\to D}(w) - \frac{m_B-m_D}{m_B+m_D} h_-^{B\to D}(w).
\end{equation}
At zero recoil ($v'=v$, so $w=1$) one expects $\mathcal{F}_{B\to D}(1)$
to be close to one, because of heavy quark symmetry.
From~(\ref{eq:differential_decay_rate}) a determination of $|V_{cb}|$
consists of the following three steps:
measure $|V_{cb}||\mathcal{F}_{B\to D}(w)|$ in an experiment,
extrapolate it to the zero-recoil limit assuming some functional form,
and use the theoretical input of $\mathcal{F}_{B\to D}(1)$.

In this paper we report on a new calculation of
$\mathcal{F}_{B\to D}(1)$ with lattice QCD,
which is model independent, at least in principle.%
\footnote{Our calculations are done in the quenched approximation,
for example, but this is a removable uncertainty and not a permanent
limitation of the method.}
The present calculation includes the leading corrections to the
heavy-quark limit: radiative corrections to the static limit
of~$h_+^{B\to D}(1)$, the $1/m_Q$ contribution of~$h_-^{B\to D}(1)$,
and the $1/m_Q^2$ contributions of~$h_\pm^{B\to D}(1)$.
Radiative corrections of order~$\alpha_s$ to~$h_-(1)$ are not yet
available, but these and further
corrections, of order $\alpha_s^2$, $\alpha_s/m_Q$, etc., could be
included in future applications of our numerical technique, once the
needed perturbative results become available.

An obvious disadvantage in using the $\bar{B}\to Dl\bar{\nu}$ decay
mode is that the branching fraction is much smaller than the
$\bar{B}\to D^*l\bar{\nu}$ mode.
Another, but not less important, shortcoming is that the phase-space
suppression factor $(w^2-1)^{3/2}$ makes the extrapolation of the
experimental data to $w=1$ more difficult than for
$\bar{B}\to D^*l\bar{\nu}$, where the corresponding factor is
$(w^2-1)^{1/2}$.
Nevertheless, the experimental result of the CLEO
Collaboration~\cite{CLEO_97} shows that the above method certainly
works, even with current statistics.
That means that future improvement of the statistics will allow a
much better determination of~$|V_{cb}|$, providing an important cross
check against other methods.

\section{HQET and the $1/m_Q$ expansion}
\label{sec:HQET_and_1/m_Q_expansion}

Many important theoretical results have been obtained for
the form factors with HQET.
The Lagrangian of HQET uses fields of infinitely heavy quarks,
so that heavy quark symmetries are manifest.
The effects of finite quark mass are included through
the $1/m_Q$ expansion and through radiative corrections.
For example, at zero recoil the form factor~$h_+$ is given by
\begin{equation}
  \label{eq:1/m_Q-expansion+}
  h_+(1) = \eta_{V} \left[ 1 - c_+^{(2)}
    \left(\frac{1}{m_c}-\frac{1}{m_b}\right)^2 + O(m_Q^{-3})
    \right],
\end{equation}
where $\eta_{V}$ represents a matching factor relating the vector
current in~(\ref{eq:definition_of_the_form_factors}) to the current in
HQET~\cite{Neubert_92a}.
The absence of the $O(1/m_Q)$ term in~(\ref{eq:1/m_Q-expansion+}) is a
result of a symmetry under an interchange of initial and final states
in~(\ref{eq:definition_of_the_form_factors}), and it is known as a
part of Luke's theorem~\cite{Luke_90}.
The same symmetry also restricts
the form of the $O(1/m_Q^2)$ terms.

The matching factor, defined so that the identity
$\mathcal{V}_0=\eta_{V} V_0^{\rm HQET}$
holds for matrix elements, is an ultraviolet- and infrared-finite
function of $m_c/m_b$.
Through one-loop perturbation theory,
\begin{equation}
  \label{eq:eta_V}
  \eta_{V^{cb}} = 1 + 3 C_F\frac{\alpha_s}{4\pi} \left(
    \frac{m_b+m_c}{m_b-m_c} \ln\frac{m_b}{m_c} -2
    \right).
\end{equation}
The two-loop coefficient is also available~\cite{Czarnecki_96}.

The vector current defined with lattice fermion fields has properties
similar to~$V_0^{\rm HQET}$.
There is a normalization factor~$Z_{V_0}$ defined so that
$\mathcal{V}_0=Z_{V_0} V_0^{\rm lat}$ holds for matrix
elements.
The factor $Z_{V_0}$ depends strongly on the (lattice) quark masses
$am_c$ and $am_b$~\cite{El-Khadra_Kronfeld_Mackenzie_97}, and its
one-loop corrections are large.
In the past, such uncertainties in the normalization prevented a
calculation of $h_+^{B\to D}(1)$ with the sought-after accuracy.
One can, however, capture most of the normalization nonperturbatively by
writing, with explicit flavor indices,
\begin{equation}
Z_{V^{cb}_0} = \sqrt{Z_{V^{cc}_0}Z_{V^{bb}_0}}\rho_{V^{cb}_0}.
\end{equation}
In our ratio~(\ref{eq:ratio_1}) the flavor-diagonal factors cancel,
so our method avoids the major normalization uncertainties.

The remaining radiative correction $\rho_{V^{cb}_0}$ depends on the
ratio of quark masses and the lattice spacing.
In the continuum limit, $am_c\to 0$ and $am_b\to 0$ with $m_c/m_b$
fixed,
\begin{equation}
\rho_{V_0}\to 1,
\end{equation}
by construction.
In the static limit, $am_c\to \infty$ and $am_b\to \infty$
with~$a$ and $m_c/m_b$ fixed,
\begin{equation}
\rho_{V_0}\to \eta_{V},
\end{equation}
because the lattice theory strictly obeys heavy-quark symmetries.
In numerical work one is somewhere in between, but the limits imply
that $\rho_{V_0}$ is never far from unity.
Two of us have computed $\rho_{V_0}$ at one loop in perturbation
theory~\cite{Kronfeld_Hashimoto_98}, verifying explicitly that the
radiative correction is small.

Similarly, the ratio~(\ref{eq:ratio_2}) is described by the expansion
\begin{equation}
  \label{eq:1/m_Q-expansion-}
  1-\frac{h_-(1)}{h_+(1)} = 1 - \beta_{V} +
    c_-^{(1)} \left(\frac{1}{m_c}-\frac{1}{m_b}\right) -
    c_-^{(2)} \left(\frac{1}{m_c^2}-\frac{1}{m_b^2}\right) +
    O(m_Q^{-3}),
\end{equation}
where $\beta_V$ is a coefficient from matching the currents
in~(\ref{eq:ratio_2}) to HQET.
Like~$\eta_V$, it is an ultraviolet- and infrared-finite function
of~$m_c/m_b$, and 
\begin{equation}
  \label{eq:beta_V}
  \beta_{V^{cb}} = 2 C_F\frac{\alpha_s}{4\pi} \left(
    \frac{2m_bm_c}{(m_b-m_c)^2} \ln\frac{m_b}{m_c}
    - \frac{m_b+m_c}{m_b-m_c}
    \right)
\end{equation}
at leading order.

The ratio~(\ref{eq:ratio_2}) again captures nonperturbatively most
of the renormalization of the lattice currents, apart from a factor
$\rho_{V^{cb}_i}$ to compensate for the difference between the radiative
corrections with a fixed lattice cutoff and with no ultraviolet cutoff.
In the continuum limit $\rho_{V_i}\to 1$, and in the static limit
$\rho_{V_i}\to 1-\beta_V$.
Again, explicit calculation verifies that the one loop contribution
remains small between the limits.

In the rest of this paper, we do not write the
matching factors $\rho_{V_\mu}$ when there is no risk of confusion.
In the final result, on the other hand, they are included.

\section{Lattice QCD and heavy quark symmetry}
\label{sec:Lattice_and_HQ}

In Ref.~\cite{El-Khadra_Kronfeld_Mackenzie_97}, it was shown that
the usual action for light quarks~\cite{Sheikholeslami_Wohlert_85}
can be analyzed in terms of the operators of HQET.
Therefore, it can be used as the basis of a systematic treatment of
heavy quarks on the lattice, even when the quark mass in lattice units,
$am_Q$, is not especially small.
The key is to adjust the couplings in the lattice action so that
operators are normalized as they are in HQET.
When $am_Q<1$, as is the case for charmed quarks at the smaller
lattice spacings in common use, this is essentially automatic, because
the higher order terms of the heavy quark expansion come from
the Dirac term of the lattice action, as in continuum QCD.
When $am_Q>1$, as is the case for bottom quarks, one can apply the
formalism of HQET to the lattice theory to obtain the normalization
conditions, as sketched below.
In either case, the kinetic energy is normalized nonperturbatively by
tuning the quark mass according to some physical condition.
Other operators are often normalized perturbatively as an initial
approximation but ultimately may be normalized nonperturbatively.

In the numerical calculations presented here, we use an action
introduced by Sheikholeslami and
Wohlert~\cite{Sheikholeslami_Wohlert_85},
\begin{equation}
    S = \sum_{x,f} \bar{\psi}_x^f\psi_x^f -
        \sum_{x,y,f} \kappa_f \bar{\psi}_x^f M_{xy}\psi_y^f +
        \frac{i}{2} c_{\rm SW} \sum_{x,f} \kappa_f
        \bar{\psi}_x^f\sigma_{\mu\nu}F_{\mu\nu}\psi_x^f,
    \label{eq:SW_action}
\end{equation}
where the index~$f$ runs over heavy and light flavors.
The hopping parameter~$\kappa_f$ is related to the bare quark mass,
\begin{equation}
    am_{0f} = \frac{1}{2\kappa_f} - \frac{1}{2\kappa_{\rm crit}},
    \label{eq:bare_mass}
\end{equation}
where~$\kappa_{\rm crit}$ is the value of~$\kappa$ needed to make a
quark massless.
The flavor-independent matrix~$M_{xy}$ vanishes except when
$y=x\pm\hat{\mu}a$, for some spacetime direction~$\mu$.
The kinetic energy arises from this term.
The gluons' field strength~$F_{\mu\nu}$ is defined on a set of paths
shaped liked a four-leaf clover, so~$S$ is often called the
``clover'' action.
With $c_{\rm SW}=0$ one has the Wilson action.

For the light quark the clover coupling~$c_{\rm SW}$ can be chosen so
that there are lattice artifacts of order~$a^2\LQCD^2$.
In our numerical work we take an approximation to the optimal value,
leaving an artifact of order $\alpha_sa\LQCD$.

For heavy quarks, the clover action~(\ref{eq:SW_action}) has the
same heavy-quark spin and flavor symmetries as continuum QCD, even
at nonzero lattice spacing.
Consequently, we can use the machinery of HQET
to characterize the lattice theory.
The same operators as in continuum QCD appear, but the coefficients
can differ.
Through first order in $1/m_Q$ there are three operators in the
heavy quark effective Hamiltonian,
\begin{equation}
    H = m_1 \bar{h}h - \frac{\bar{h}\vek{D}^2h}{2m_2} -
    i\frac{\bar{h}\vek{\Sigma}\cdot\vek{B}h}{2m_B} + \cdots ,
    \label{eq:H}
\end{equation}
where~$h$ is a heavy quark field, and the coefficients $m_1$, $1/m_2$,
and $1/m_B$ depend on the bare mass and the gauge coupling.
Because the lattice breaks relativistic invariance, the three
``masses'' are not necessarily equal, except as $am_0\to 0$.

At tree level, the rest mass $am_1=\log(1+am_0)$, and the
(inverse) kinetic mass
\begin{equation}
    \frac{1}{am_2}=\frac{2}{am_0(2+am_0)} + \frac{1}{1+am_0}.
    \label{eq:m2}
\end{equation}
The first term can be traced to the Dirac term of the lattice
action, and the second to the Wilson term.
The one-loop corrections to $am_1$ and $am_2$ are also
available~\cite{Mertens_Kronfeld_El-Khadra_98}.
The chromomagnetic mass~$m_B$ is considered below.

In the heavy quark effective theory, the rest mass term~$m_1\bar{h}h$
commutes with the rest of the Hamiltonian and, thus, decouples from the
dynamics.
As with decay constants~\cite{Kronfeld_95}, one can
derive the expansions like~(\ref{eq:1/m_Q-expansion+})
and~(\ref{eq:1/m_Q-expansion-}) within the lattice theory, and the
rest mass disappears from physical amplitudes~\cite{Kronfeld_99}.
On the other hand, adjusting the bare quark mass so that $m_2=m_Q$ is
the way to normalize the kinetic operator ${\bar{h}\vek{D}^2h}/{2m_2}$
correctly.
This normalization can be implemented nonperturbatively by demanding
that the energy of a hadron have the correct momentum dependence.
In our numerical work we use the $B$ and $D$ mesons for this
purpose.
Furthermore, one can correctly normalize the
chromomagnetic operator~${\bar{h}\vek{\Sigma}\cdot\vek{B}h}/{2m_B}$
by adjusting the clover coupling~$c_{\rm SW}$, as a
function of the gauge coupling, so that $m_B=m_2$.
For example, at tree level the desired adjustment is $c_{\rm SW}=1$.
In our numerical work, we choose $c_{\rm SW}$ in a way that sums up
tadpole diagrams, which dominate perturbation theory.
This amounts to normalizing the chromomagnetic operator perturbatively.

In summary, we adjust the bare mass~$am_0$ and clover
coupling~$c_{\rm SW}$ so that the leading effects of the heavy-quark
expansion are correctly accounted
for~\cite{El-Khadra_Kronfeld_Mackenzie_97}.
Previous work in the literature chose instead to adjust the bare mass
until $m_1=m_Q$, which introduces an unnecessarily large error,%
\footnote{To mitigate this error, these calculations are often carried
out at artificially small quark masses.
Ensuing extrapolations to larger masses contaminate lower orders in
the (physical) $1/m_Q$ expansion with higher orders.}
proportional to $1-m_1(m_0)/m_2(m_0)$.

Under renormalization the heavy quark kinetic energy can mix with the
rest mass term in a power divergent way.
Because the lattice action used here contains both, the rest mass
fully absorbs the power divergence.
A related problem is the ambiguity owing to
renormalons~\cite{Beneke_98},
which appears in some quantities in HQET or nonrelativistic QCD
(NRQCD).
It is irrelevant to our work, because we calculate physical quantities,
namely the masses of the $B$ and $D$ mesons and decay amplitude for
$\bar{B}\to Dl\bar{\nu}$.

To complete the correspondence of the lattice theory to HQET we must
consider the vector current.
At order $1/m_Q$ of HQET
\begin{equation}
  V_\mu^{cb} =
  \left(\bar{h}^c+\vek{D}\bar{h}^c\cdot\frac{\veg{\gamma}}{2m_{3c}}\right)
  \gamma_\mu
  \left(1-\frac{\veg{\gamma}\cdot\vek{D}}{2m_{3b}}\right)h^b +\cdots,
  \label{eq:Vhh}
\end{equation}
where the coefficient $1/m_3$ depends on the current employed.
The heavy-heavy current on the lattice is constructed by defining a
rotated field~\cite{El-Khadra_Kronfeld_Mackenzie_97,Kronfeld_95},
\begin{equation}
\Psi^f = \sqrt{2\kappa_f}
\left[1+ad_1^f(am_{0f},g_0^{2})\veg{\gamma}\cdot\vek{D}\right]\psi^f,
\label{eq:rotation}
\end{equation}
where $\psi$ is the quark field in the hopping-parameter form of the
action~(\ref{eq:SW_action}).
Then the lattice vector current
\begin{equation}
    V^{cb}_\mu = \bar{\Psi}^c \gamma_\mu \Psi^b
    \label{eq:V_0}
\end{equation}
and $\mathcal{V}^{cb}_\mu=Z_{V^{cb}_\mu}V^{cb}_\mu$.
Both~$Z_{V^{fg}_\mu}$ and~$d_1^f$ depend on the gauge coupling, the
masses, and (at higher orders) on the Dirac matrix in~(\ref{eq:Vhh}).
They are adjusted so that the normalization and momentum dependence of
matrix elements matches the continuum, respectively.
In particular, at tree level the coefficient in~(\ref{eq:Vhh}) is
\begin{equation}
  \frac{1}{am_3} = \frac{2(1+am_0)}{am_0(2+am_0)} - 2d_1,
  \label{eq:m3}
\end{equation}
and the condition $m_3=m_2$ prescribes a condition
on~$d_1$~\cite{El-Khadra_Kronfeld_Mackenzie_97,Kronfeld_95}.

From the properties of the operators under heavy-quark symmetry,
it follows that the~$1/m_2$ and~$1/m_B$ terms in~(\ref{eq:H}) could
give a contribution to~$h_+(1)$, but not to~$h_-(1)$~\cite{Luke_90}.
On the one hand, these contributions to~$h_+(1)$ must be symmetric
under interchange of the initial and final states, but, on the other
hand, they must vanish when the initial and final quark masses are
the same.
Consequently, there can be no contributions linear in either~$1/m_2$
or~$1/m_B$.
Our definition of~$h_+(1)$ enjoys this property, by construction,
because~(\ref{eq:ratio_1}) manifestly respects the interchange
symmetry.

Similarly, the $1/m_3$ terms in~(\ref{eq:Vhh}) give a contribution
only to~$h_-(1)$.
It must be anti-symmetric under interchange of the initial and final
states and must vanish when the initial and final quark masses are
the same.
Our definition of~$h_-(1)$, again by construction, ensures that
only the combination $1/m_{3c}-1/m_{3b}$ appears.
This feature is taken into account in
Sec.~\ref{sec:Heavy_quark_mass_dependence_of_h-}.

In~(\ref{eq:1/m_Q-expansion+}) and~(\ref{eq:1/m_Q-expansion-})
we seek contributions of order~$1/m_Q^2$.
These come from double insertions of the $1/m_Q$ terms
in~(\ref{eq:H}) and~(\ref{eq:Vhh}), and from $1/m_Q^2$ terms implied
by the ellipses.
Remarkably, the latter cancel when~$h_+(1)$ and~$h_-(1)$ are defined by
the double ratios~(\ref{eq:ratio_1})
and~(\ref{eq:ratio_2})~\cite{Kronfeld_99}.
This is easy to understand if one starts with the matrix elements.
The $1/m_b^2$ and $1/m_c^2$ corrections to the action arise from the
initial or final state only.
To this order, one can factorize them.
They drop out of the double ratios, because the numerator and
denominator of~(\ref{eq:ratio_1}) or~(\ref{eq:ratio_2}) contain the
same number of $B$ and $D$ factors.
The same applies to $1/m_b^2$ and $1/m_c^2$ corrections to the current.
There may be a nonfactorizable correction to the current with
coefficient $C(m_c,m_b)/m_cm_b$, where the function $C$ is unknown,
except that in perturbation theory it starts at one loop
and that $C(m,m)=0$.

In the long run, one would like to pick up terms of order~$1/m_Q^3$
and higher.
Because the bottom quark is so heavy, these are dominated by the
$1/m_c^n$ terms.
With the normalization conditions outlined
here~\cite{El-Khadra_Kronfeld_Mackenzie_97}, these come automatically
from the Dirac term, as in continuum QCD.
In future work, at smaller lattice spacings, the Dirac term will
dominate, generating contributions to all orders in~$1/m_c$.

\section{Lattice details}
\label{sec:Lattice_details}

Our numerical data are obtained in the quenched approximation
on a $12^3\times 24$ lattice with the plaquette gluon action
at~$\beta=6/g_0^2=5.7$.
We take a mean-field-improved~\cite{Lepage_Mackenzie_93} value of the
clover coupling, which on this lattice is $c_{\rm SW}=1.57$.
Out of 300 configurations generated for our previous
work~\cite{Fermilab_97}, we use 200 configurations.
We usually define the inverse lattice spacing through the charmonium
1S--1P splitting, finding $a^{-1}(\mbox{1S--1P})=1.16^{+3}_{-3}$~GeV.
For comparison, with the kaon decay constant
$a^{-1}(f_K)=1.01^{+2}_{-1}$~GeV, and the difference is thought to be
part of the error of quenching.
Because the form factors are dimensionless, the lattice spacing
affects them only indirectly, through the adjustment of the quark
masses.

To investigate the heavy quark mass dependence of the form factors
we take $\kappa_h=0.062$, 0.089, 0.100, 0.110, 0.119 and 0.125, and
consider several combinations for the heavy quarks in the initial
and final states.
The mass of the spectator light quark is usually taken to be close
to that of the strange quark, for which $\kappa_l=0.1405$.
We examine the effect of chiral extrapolation using four
$\kappa_l$ values, 0.1405, 0.1410, 0.1415, and 0.1419, for various
combinations of the initial and final heavy quark masses
$\kappa_h= 0.089$, 0.110, and 0.119.
The critical hopping parameter is
$\kappa_{\rm crit}=0.14327^{+5}_{-3}$.

For the computation of the matrix element%
\footnote{For simplicity we use ``$B$'' instead of ``$\bar{B}$'' to
indicate the $(b\bar{q})$ meson, and we use ``$B$'' or
``$D$'' for any values of the heavy quark masses.}
$\langle D(\vek{p}')|V_{\mu}|B(\vek{p})\rangle$
we calculate the three point correlation function
\begin{equation}
  \label{eq:three_point_correlator}
  C^{DV_{\mu}B}(t,\vek{p}',\vek{p}) =
  \sum_{\vek{y},\vek{x}}
  e^{-i(\vek{p}-\vek{p}') \cdot \vek{y}}
  e^{-i\vek{p} \cdot \vek{x}}
  \langle D(0,\mathbf{0}) V_{\mu}(t,\vek{y})
  B^{\dagger}(T/2,\vek{x}) \rangle
\end{equation}
with $V_\mu$ from~(\ref{eq:V_0}) and $\vek{p}=\mathbf{0}$.
The light quark propagator is solved with a source at time
slice $0$, and we place the interpolating field for $B$ at
$T/2$, where we use the source method.
The interpolating fields $B$ and $D$ are constructed with
the 1S state smeared source as in Ref.~\cite{Fermilab_97}.
The spatial momentum $\vek{p}'$ carried by the final
state is taken to be (0,0,0), (1,0,0), (1,1,0), (1,1,1) and
(2,0,0) in units of $2\pi/L$, where $L$ is the physical size of
the box; in our case, $L=12a$.

The numerical results presented below are obtained from uncorrelated
fits to ratios of these three-point functions.
The statistical errors are estimated with the jackknife method.
For a subset of the data we have repeated the analysis with
correlated fits and the bootstrap method.
We find no statistically significant difference.

In much of the numerical work presented in this paper,
we set the coefficients~$d_1$ of the
rotation~(\ref{eq:rotation}) to zero.
From the discussion following~(\ref{eq:Vhh}) the dependence on~$d_1$
enters directly through~$1/m_3$, and indirectly by
changing~$\rho_{V_\mu}$.
On the scattering matrix elements of the spatial current~$V_i$,
this should make a small ($\lsim 10~\%$ or so) effect.
On the temporal current~$V_0$, the effect should be tiny.
Both expectations are checked at representative choices of the heavy
quark masses, and the uncertainty introduced into the spatial current
is propagated to the final result.

\section{Calculation of $|h_+(1)|$}
\label{sec:Calculation_of_h_+}

The form factor $|h_+(w)|$ at zero recoil is obtained directly from
the three-point correlation
functions~(\ref{eq:three_point_correlator}), setting all three
momentum to be zero.
We define a ratio%
\footnote{Mandula and Ogilvie~\cite{Mandula_Ogilvie_94} used a similar
ratio, with nonzero velocity transfer, to study the $w$ dependence of
the Isgur-Wise function, which is the infinite mass limit of~$h_+(w)$.}
\begin{equation}
  \label{eq:Ratio_for_h+}
  R^{B\to D}(t) \equiv
  \frac{C^{DV_{0}B}(t, \mathbf{0},\mathbf{0})
        C^{BV_{0}D}(t, \mathbf{0},\mathbf{0})}
       {C^{DV_{0}D}(t, \mathbf{0},\mathbf{0})
        C^{BV_{0}B}(t, \mathbf{0},\mathbf{0})},
\end{equation}
in which the exponential dependence on $t$ associated with
the ground state masses cancels between the numerator and
denominator.
When the current and two interpolating fields
are separated far enough from each other, the contribution of the
ground state dominates and
\begin{eqnarray}
  \label{eq:Ratio_for_h+_limit}
  R^{B\to D}(t) & \to &
  \frac{\langle D(\mathbf{0})|V_0|B(\mathbf{0})\rangle
        \langle B(\mathbf{0})|V_0|D(\mathbf{0})\rangle}
       {\langle D(\mathbf{0})|V_0|D(\mathbf{0})\rangle
        \langle B(\mathbf{0})|V_0|B(\mathbf{0})\rangle}
      \nonumber \\
  & = &
  \frac{|h_+^{B\to D}(1) h_+^{D\rightarrow B}(1)|}
       {|h_+^{D\to D}(1) h_+^{B\rightarrow B}(1)|}
  = |h_+^{B\to D}(1)|^2,
\end{eqnarray}
suppressing radiative corrections.
Here we use the definition~(\ref{eq:definition_of_the_form_factors})
and the unit normalization of $|h_+(1)|$ in the equal mass case.
Thus, we expect $R^{B\to D}$ to be constant as a function of~$t$,
and its value represents the form factor squared.

\begin{figure}[btp]
  \begin{center}
    \epsfxsize=12cm \epsfbox{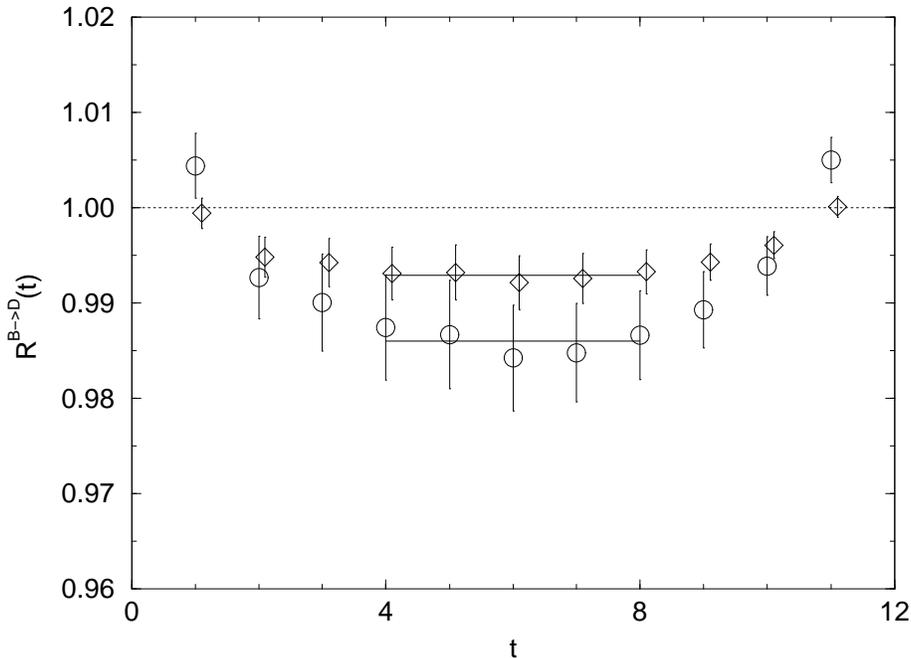}
    \caption{$R^{B\to D}(t)$ as a function of $t$.
      The heavy quark hopping parameter for the initial and
      final mesons are $(\kappa_b,\kappa_c) =
      (0.089,0.110)$ (diamonds), and $(0.089,0.119)$
      (circles).
      The light quark corresponds to the strange quark,
      $\kappa_l=0.1405$.
      The solid lines represent a constant fit with
      $4\leq t\leq 8$.}
    \label{fig:R_BtoD}
  \end{center}
\end{figure}

In Fig.~\ref{fig:R_BtoD} we plot the ratio $R^{B\to D}(t)$ for
two representative combinations of mass parameters.
We observe a nice plateau extending over about five time slices, and
our fit over the interval $4\leq t\leq 8$ is shown by the solid line.

\begin{figure}[btp]
  \begin{center}
    \epsfxsize=12cm \epsfbox{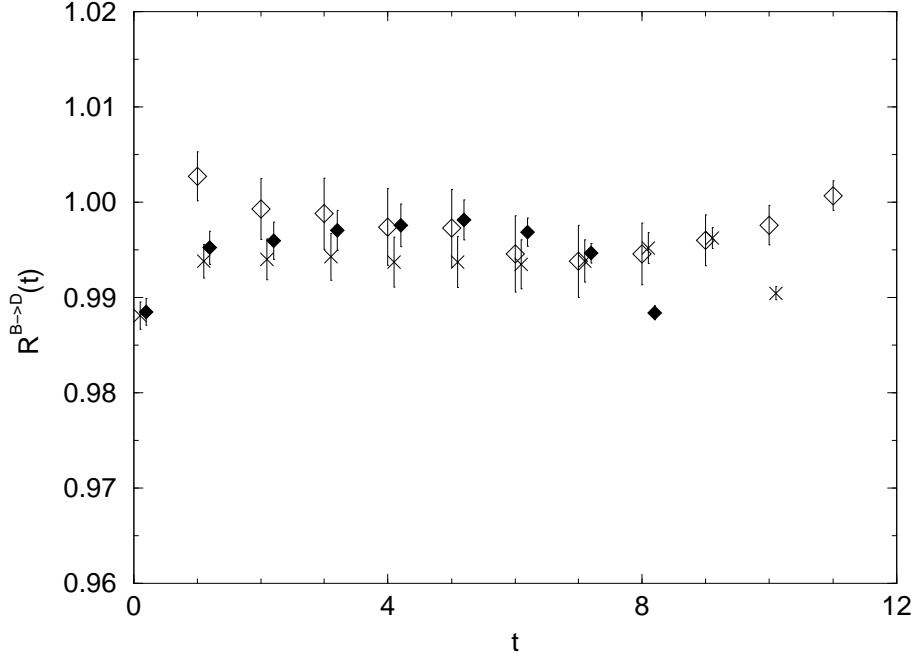}
    \caption{Check of the plateau in $R^{B\to D}(t)$
      by varying the time slice $t_B$ of the $B$ meson
      interpolating field.
      Open diamonds, crosses and solid diamonds correspond
      to the results with $t_B=12$, 10, and 8, respectively.
      The heavy quark hopping parameters are
      $(\kappa_b,\kappa_c) = (0.089,0.110)$, and
      $\kappa_l=0.1405$.}
    \label{fig:R_BtoD_tB}
  \end{center}
\end{figure}

To see if the plateau is stable under the change of the position of the
interpolating field, we repeat the calculation changing the time~$t_B$
of the $B$-meson interpolating field.
The results with $t_B=10$ and 8 are shown in
Fig.~\ref{fig:R_BtoD_tB} together with the one with $t_B=T/2=12$.
We observe that the plateau is very stable and conclude that the
extraction of the ground state is reliable.
In the following analysis we use the result with $t_B=T/2$,
and the numerical data for each $\kappa_h$ are given in
Table~\ref{tab:h+_data}.
\begin{table}[tbp]
  \begin{center}
    \caption{Numerical data for $R^{B\to D}$, which
      corresponds to $|h_+(1)|^2$, at $\kappa_l=0.1405$.
      Rows (columns) are labeled by the value of $\kappa_h$ in the
      initial (final) state.
      Combinations without data have not been calculated in this work.
      The diagonal elements are~1 by construction.}
    \label{tab:h+_data}
\vskip0.3em
    \begin{tabular}{l|llllll}
\hline\hline
\multicolumn{1}{c|}{$\kappa_h$}
      & \multicolumn{1}{c}{0.062} & \multicolumn{1}{c}{0.089} &
        \multicolumn{1}{c}{0.100} & \multicolumn{1}{c}{0.110} &
        \multicolumn{1}{c}{0.119} & \multicolumn{1}{c}{0.125} \\
\hline
0.062 &1        &0.989(07)&0.979(12)&         &         &0.947(24)\\
0.089 &0.989(07)& 1       &0.998(01)&0.993(02)&0.986(05)&0.983(07)\\
0.100 &0.979(12)&0.998(01)&1        &         &         &0.992(04)\\
0.110 &         &0.993(02)&         &1        &0.999(01)&         \\
0.119 &         &0.986(05)&         &0.999(01)&1        &         \\
0.125 &0.947(24)&0.983(07)&0.992(04)&         &         &1        \\
\hline\hline
    \end{tabular}
  \end{center}
\end{table}

\begin{figure}[btp]
  \begin{center}
    \epsfxsize=12cm \epsfbox{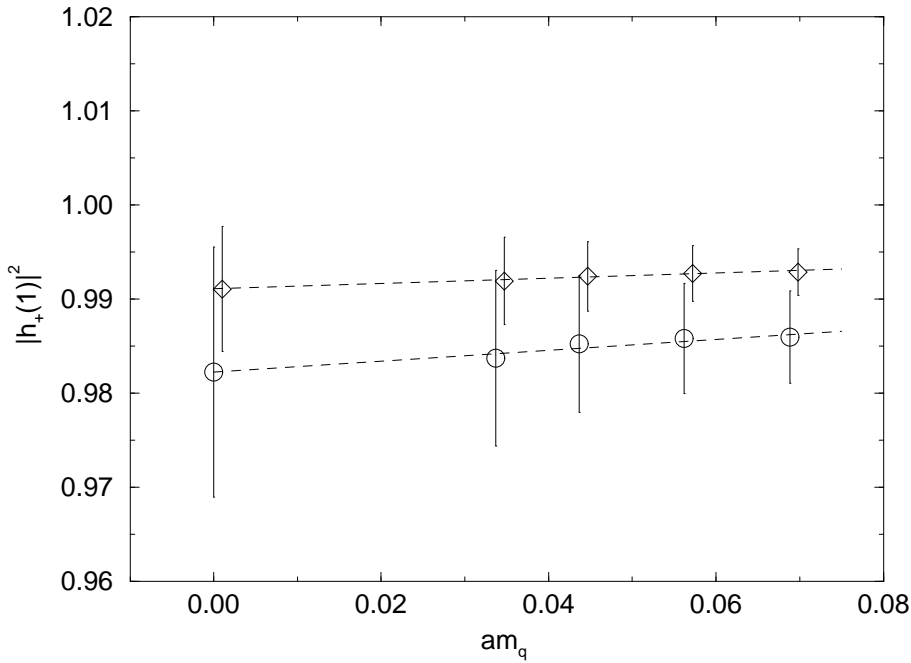}
    \caption{Chiral extrapolation of $|h_+(1)|^2$.
      The heavy quark hopping parameters for the initial and
      final mesons are $(\kappa_b,\kappa_c) = (0.089,0.110)$
      (diamonds), and $(0.089,0.119)$ (circles).
      }
    \label{fig:h+_chiral}
  \end{center}
\end{figure}

We examine the chiral limit by computing with four values of the
light quark mass~(\ref{eq:bare_mass}), roughly in the range
$m_s/2\leq m_q \leq m_s$.
Figure~\ref{fig:h+_chiral} shows that the $am_q$ dependence of
$|h_+(1)|^2$, for two combinations of~$(\kappa_b,\kappa_c)$, is
very slight.
A linear fit in $am_q$ gives a slope consistent with zero, and the
value in the chiral limit is still consistent with that at the finite
light quark mass.
With our present statistics, we cannot study the dependence on the
light and heavy quark masses simultaneously.
Instead we take from Fig.~\ref{fig:h+_chiral} two lessons:
the dependence on the light quark mass is insignificant,
but the (statistical) uncertainty increases, by a factor of two,
in the chiral limit.

A small, but non-analytic, dependence on $m_{\pi}$ is expected from
chiral perturbation theory~\cite{Randall_Wise_93,Boyd_Grinstein_95}.
Such effects may be different in the quenched approximation.
If so, the difference should be counted as part of the error of the
quenched approximation.

\section{Heavy quark mass dependence of $|h_{+}(1)|$}
\label{sec:Heavy_quark_mass_dependence_of_h+}

In the heavy quark limit of QCD, the heavy quark mass dependence
of~$|h_+(1)|$ can be described with a $1/m_Q$ expansion.
Using a symmetry of its
definition~(\ref{eq:definition_of_the_form_factors}) under the exchange
of the initial and final states and the normalization in the limit
of degenerate heavy quark mass, the form of the $1/m_b$ and $1/m_c$
expansion is restricted to be
\begin{equation}
\label{eq:1/m-expansion_of_h+}
  |h_+(1)| = 1 -
    c_+^{(2)} \left(\frac{1}{m_c}-\frac{1}{m_b}\right)^2 +
    c_+^{(3)} \left(\frac{1}{m_c}+\frac{1}{m_b}\right)
      \left(\frac{1}{m_c}-\frac{1}{m_b}\right)^2         + O(1/m_Q^4),
\end{equation}
suppressing the radiative correction~$\eta_{V}$.
The term $O(1/m_Q^4)$ denotes all possible combinations of
$1/m_c$ and $1/m_b$ with total mass dimension~$-4$.
The absence of terms of order $1/m_Q$
is implied by Luke's theorem~\cite{Luke_90}, but in this
particular case it can be understood as a result of the
symmetry $1/m_c \leftrightarrow 1/m_b$.

If we take the radiative corrections into account, the data presented in
the last section correspond to $|h_+(1)|/\rho_{V_0}$.
To use the right-hand side of~(\ref{eq:1/m-expansion_of_h+}), on the
other hand, we must multiply by with $\rho_{V_0}/\eta_{V}$ to
obtain $|h_+(1)|/\eta_{V}$.
At $\beta=5.7$ and our choices of quark masses we find, at one loop,
that $\rho_{V_0}/\eta_{V}$ is very nearly~1, so that we do not
need to carry out this conversion.%
\footnote{This is an accident at our choice of lattice spacing.
For smaller lattice spacings, this would not be so.
See Ref.~\cite{Kronfeld_Hashimoto_98} for details.}

In the lattice theory, the masses
entering~(\ref{eq:1/m-expansion_of_h+}) are $m_2$, $m_B$, and $m_3$,
as explained in Sec.~\ref{sec:Lattice_details}.
In particular, if one follows Refs.~\cite{Falk_Neubert_93,Mannel_94}
to see how higher-dimension tree-level operators affect the matrix
elements, one sees that the~$1/m_c^2$ and~$1/m_b^2$ corrections to
the action and current do not affect~$R^{B\to D}$~\cite{Kronfeld_99}.

We study the relation~(\ref{eq:1/m-expansion_of_h+})
with several combinations of the initial and final heavy
quark masses.
We require a relation between the hopping parameters, which are inputs
to the numerical calculation, and the quark masses.
To simplify the analysis, we set $m_3=m_B=m_2$ and estimate the kinetic
quark mass by applying tadpole improvement~\cite{Lepage_Mackenzie_93}
to include the dominant tadpole contribution to the perturbation
series.
The tadpole-improved kinetic mass is given by substituting
$a\tilde{m}_0=am_0/u_0$ for~$am_0$ on the right-hand side
of~(\ref{eq:m2}), with the mean link variable
$u_0=1/8\kappa_{\rm crit}$.
We do not bother with the one-loop correction
to~$m_2$~\cite{Mertens_Kronfeld_El-Khadra_98},
because it is smaller than the uncertainty from~$a$.
This way of parametrizing the quark masses is for interpolating
only; when reconstituting the physical result, the hopping
parameters~$\kappa_c$ and~$\kappa_b$ are chosen nonperturbatively
from the masses of the~$D$ and~$B$ mesons.

\begin{figure}[btp]
  \begin{center}
    \epsfxsize=12cm \epsfbox{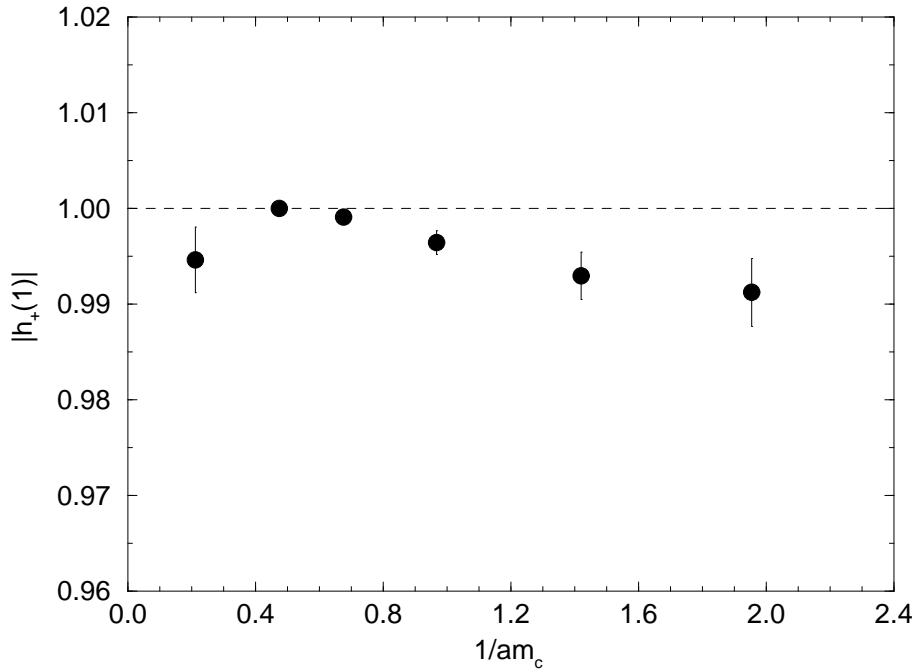}
    \caption{$1/am_c$ dependence of $|h_+(1)|$.
      The initial heavy quark mass is fixed at
      $\kappa_b=0.089$, which corresponds to $1/am_b=0.475$.
      The light quark corresponds to the strange quark,
      $\kappa_l=0.1405$.}
    \label{fig:h+_heavy_mass_dependence}
  \end{center}
\end{figure}

Figure~\ref{fig:h+_heavy_mass_dependence} shows the $1/am_c$
dependence of~$|h_+(1)|$.
The initial heavy quark mass is set to be $1/am_b=0.475$
($\kappa_b=0.089$), and we vary $1/am_c$ between~$0.2$ and~$2.0$.
(Here we misuse the meaning of subscript $b$ or $c$ to
indicate the initial or final state heavy quark,
respectively.)
At $1/am_c=1/am_b$ the form factor becomes exactly one by
construction, and the deviation from unity increases as $1/am_c$
moves away from $1/am_b$.
The statistical error grows as the difference of heavy quark
masses increases.
When one approaches the static limit the signal becomes
much noisier, as in many other Monte Carlo calculations with
heavy-light mesons.
In our case, the statistical error of the point with heaviest~$am_c$
is very large.

\begin{figure}[btp]
  \begin{center}
    \epsfxsize=12cm \epsfbox{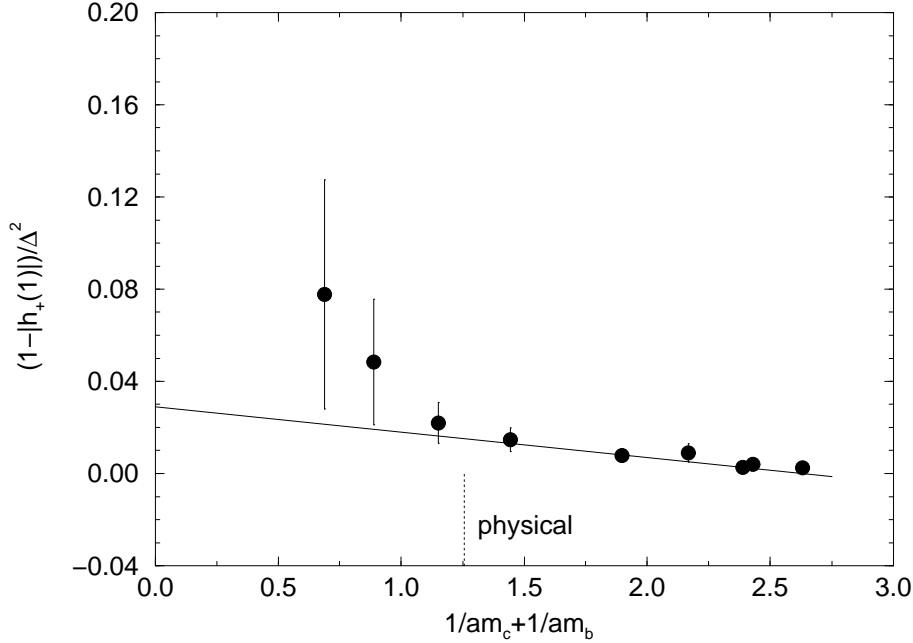}
    \caption{
      $\left[1-|h_+(1)|\right]/\Delta^2$ vs.\ $1/am_c+1/am_b$.
      The dotted vertical line indicates the physical value of
      $1/am_c+1/am_b$.
      The light quark corresponds to the strange quark,
      $\kappa_l=0.1405$.}
    \label{fig:h+_heavy_mass_coeff}
  \end{center}
\end{figure}

To see the mass dependence more clearly, we rewrite the
relation~(\ref{eq:1/m-expansion_of_h+}) as
\begin{equation}
  \label{eq:1/m-dependence_of_h+_reduced}
  \frac{1-|h_+(1)|}{\Delta^2} = c_+^{(2)} - c_+^{(3)}
  \left( \frac{1}{am_c} + \frac{1}{am_b} \right),
\end{equation}
where $\Delta = 1/am_c - 1/am_b$.
The left-hand side is plotted in Fig.~\ref{fig:h+_heavy_mass_coeff}.
The data exhibit a very good linear dependence on $\Delta$,
except in the heavy mass regime, where the error grows rapidly.
Fitting all data linearly, we obtain
$c_+^{(2)}=0.029(11)$ and $c_+^{(3)}=0.011(4)$.
In  physical units, and absorbing factors of $a$ into the coefficients,
these coefficients have a size typical of the QCD scale:
$c_+^{(2)}=[0.20(4)~\mathrm{GeV}]^2$ and
$c_+^{(3)}=[0.26(3)~\mathrm{GeV}]^3$.

The dotted line marking the physical value of $1/am_c+1/am_b$ shows
that we are, in effect, using~(\ref{eq:1/m-dependence_of_h+_reduced})
as an Ansatz for interpolation.
Although the coefficients are interesting in their own right, we
caution the reader that the values extracted from the fit are highly
correlated, and we have not made a full analysis of the errors on them.
Below we prefer to give~$h_+(1)$, evaluated at physical values of the
masses, as the principal result of this section.

We have checked the influence of the rotation by repeating the
calculations with~$d_1$ set to
\begin{equation}
    \label{eq:tilde_d_1}
    \tilde{d}_1 =
      \frac{1}{2+a\tilde{m}_0} - \frac{1}{2(1+a\tilde{m}_0)}
\end{equation}
at several of the heavy-quark mass combinations.
This is the correctly tuned value at (mean-field improved) tree
level~\cite{El-Khadra_Kronfeld_Mackenzie_97}.
The primary effect of varying~$d_1$ is through the combination
$(1/m_{3c}-1/m_{3b})^2$~\cite{Kronfeld_99}.
A secondary effect is to modify the radiative corrections
factor~$\rho_{V_0}$.
After interpolating the masses to the physical point, the change
on~$h_+(1)$ is~$+0.00013$, which is much smaller than several other
uncertainties.
Owing to this, our central value for~$h_+(1)$ can come safely from
data with $d_1=0$, the only value of~$d_1$ for which~$\rho_{V_0}$
is already available.

\section{Calculation of $h_-(1)$}
\label{sec:Calculation_of_h_-}

To obtain $h_-(w)$, it is necessary to consider nonzero recoil
momentum.
From the definition of the form
factors~(\ref{eq:definition_of_the_form_factors}), the matrix elements
of the spatial and temporal vector current for the nonzero recoil
final state $D(\vek{p}')$ read
\begin{equation}
  \label{eq:V_i_and_V_0_matrix elements}
  \langle D(\vek{p}') |\mathcal{V}_i| B(\mathbf{0}) \rangle=
  \sqrt{m_B m_D} \left[ h_+^{B\to D}(w) -
    h_-^{B\to D}(w) \right] v'_i,
\end{equation}
\begin{equation}
  \langle D(\vek{p}') |\mathcal{V}_0| B(\mathbf{0}) \rangle=
  \sqrt{m_B m_D} \left[ h_+^{B\to D}(w) (1+w)
    + h_-^{B\to D}(w) (1-w) \right],
\end{equation}
where $w=v\cdot v'=\sqrt{1+\vek{v}'^2}$,
and $\vek{v}'= \vek{p}'/m_D$.

On the lattice we start by computing the ratio of correlation functions
\begin{equation}
  \label{eq:ratio_for_h-}
  R^{B\to D}_{V_i/V_0}(t,\vek{p}') \equiv
  \frac{C^{DV_{i}B}(t,\vek{p}',\mathbf{0})}
       {C^{DV_{0}B}(t,\vek{p}',\mathbf{0})}.
\end{equation}
In the limit of well-separated currents, the time dependence flattens,
\begin{eqnarray}
  \label{eq:ratio_for_h-_limit}
  R^{B\to D}_{V_i/V_0}(t,\vek{p}') & \to &
      \frac{\langle D(\vek{p}')|V_i|B(\mathbf{0})\rangle}
           {\langle D(\vek{p}')|V_0|B(\mathbf{0})\rangle} \\
  & = & \frac{v'_i}{2} \left[1-
  \frac{h^{B\to D}_-(w)}{h^{B\rightarrow D}_+(w)}\right]
  \left[ 1 - \frac{1}{2}\left(1-
    \frac{h^{B\to D}_-(w)}{h^{B\rightarrow D}_+(w)}
    \right)(w-1)
  \right]. \nonumber
\end{eqnarray}
The last step holds for small~$\vek{v}'^2$ and
suppresses radiative corrections.
Because the velocity inherits statistical uncertainties from the $D$'s
kinetic mass, it is further useful to define a double ratio
\begin{equation}
  R^{(B\to D)/(D\to D)}_{V_i/V_0}(t,\vek{p}') \equiv
    R^{B\to D}_{V_i/V_0}(t,\vek{p}') /
    R^{D\to D}_{V_i/V_0}(t,\vek{p}') .
\end{equation}
Then, for large time separations,
\begin{eqnarray}
  \label{eq:ratio_of_ratio_for_h-}
  R^{(B\to D)/(D\to D)}_{V_i/V_0}(t,\vek{p}') & \to &
      \frac{\langle D(\vek{p}')|V_i|B(\mathbf{0})\rangle}
           {\langle D(\vek{p}')|V_0|B(\mathbf{0})\rangle}
      \frac{\langle D(\vek{p}')|V_0|D(\mathbf{0})\rangle}
           {\langle D(\vek{p}')|V_i|D(\mathbf{0})\rangle} \\
  & = &
  \left[ 1 - \frac{h^{B\to D}_-(w)}{
                   h^{B\to D}_+(w)} \right]
  \left[ 1 + \frac{h^{B\to D}_-(w)}{2
                   h^{B\to D}_+(w)}
             (w-1) \right]. \nonumber
\end{eqnarray}
The final expression is simplified using the property
$h^{D\to D}_-(w)=0$.
Provided that $|h^{B\to D}_+(w)|$ is obtained sufficiently precisely in
the previous sections, the relation~(\ref{eq:ratio_of_ratio_for_h-})
can be used to extract $h^{B\to D}_-(w)$.
The part proportional to $w-1$ gives
only a small contribution, since the coefficient
$h_-(w)/h_+(w)$ is itself a small quantity of order
$(m_B-m_D)/(m_B+m_D)$.

\begin{figure}[btp]
  \begin{center}
    \epsfxsize=12cm \epsfbox{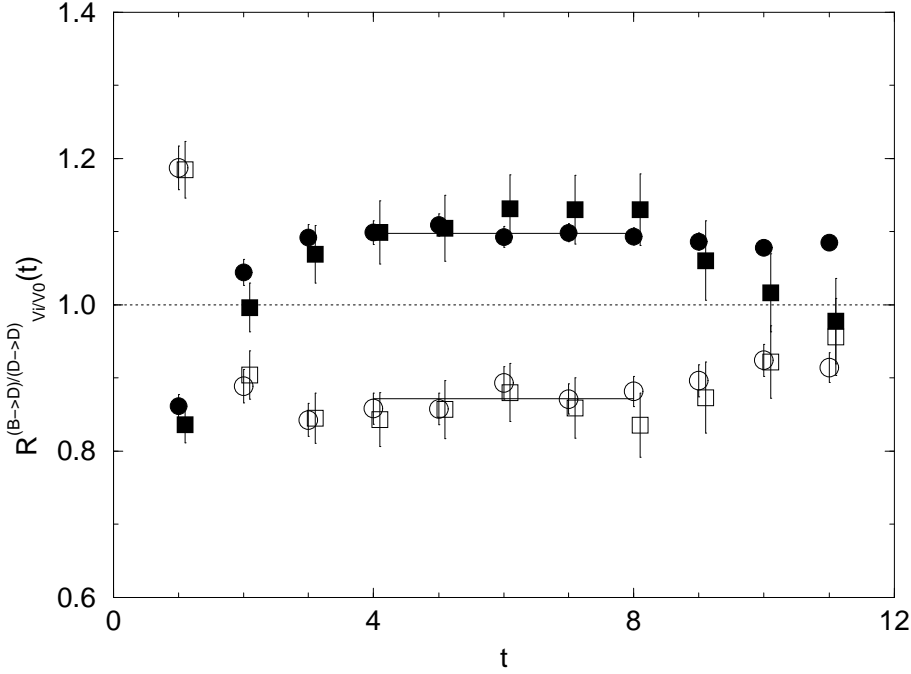}
    \caption{
      $R^{(B\to D)/(D\rightarrow D)}_{V_i/V_0}(t)$
      for the final state momentum $(1,0,0)$ (circles) and
      $(2,0,0)$ (squares).
      The heavy quark hopping parameter for the initial and
      final mesons are $(\kappa_b,\kappa_c) = (0.089,0.119)$
      (solid symbols) and $(0.119,0.089)$ (open symbols).
      The light quark corresponds to the strange quark,
      $\kappa_l=0.1405$.
      The solid lines represent a constant fit for the
      momentum $(1,0,0)$ with $4\leq t\leq 8$.}\label{fig:R_BDoverDD}
  \end{center}
\end{figure}

Figure~\ref{fig:R_BDoverDD} shows the $t$ dependence of the
ratio $R^{(B\to D)/(D\rightarrow D)}_{V_i/V_0}(t,\vek{p}')$
for  final state momenta
$L\vek{p}'/2\pi=(1,0,0)$ (circles) and $(2,0,0)$ (squares).
Filled symbols represent the $b\to c$ transition, while
open symbols correspond to the reverse $c\to b$
transition.
The plateau is reached around $t=4$, so that we can fit in the
interval $4\leq t\leq 8$, as with~$|h_+(1)|^2$.
The fit results for the momentum $(1,0,0)$ are given by the
solid lines.

Up to the small contribution of order $w-1$, this ratio gives
the combination $1-h^{B\to D}_-(w)/h^{B\rightarrow D}_+(w)$, in which
$h^{B\to D}_+(w)$ is almost equal to~1.
Looking at the solid symbols in Fig.~\ref{fig:R_BDoverDD},
$h^{B\to D}_-(w)$ is roughly~$-0.1$ and is almost independent of the
final state momentum.
Since $h_-(w)$ changes its sign under the exchange of
initial and final states, it is consistent that the open
symbols, which correspond to the transition $D\to B$,
appear below one.

\begin{figure}[btp]
  \begin{center}
    \epsfxsize=12cm \epsfbox{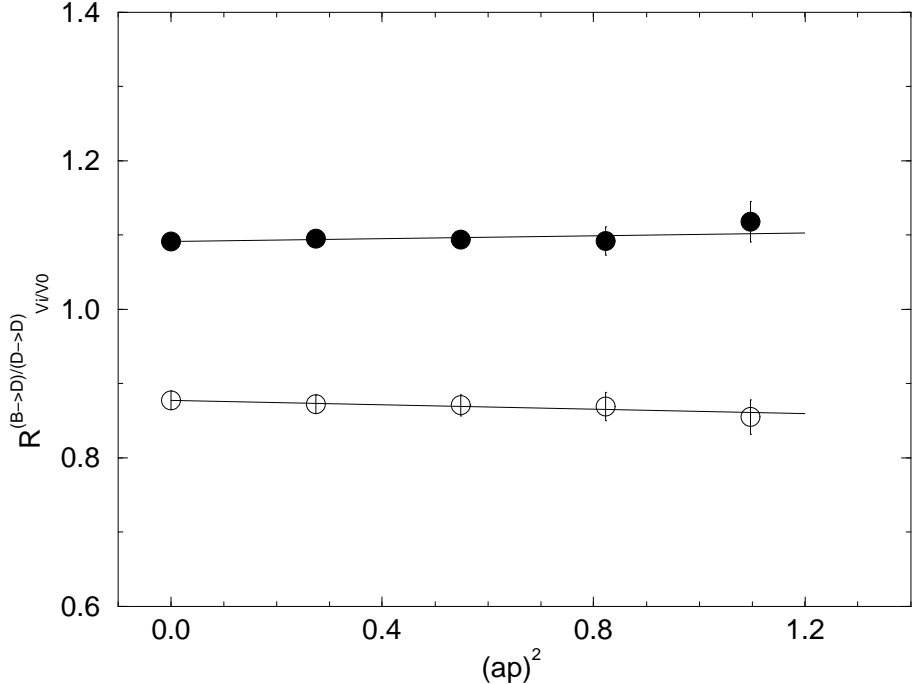}
    \caption{
      Extrapolation of
      $R^{(B\to D)/(D\to D)}_{V_i/V_0}$ to the
      zero-recoil limit.
      The heavy quark hopping parameter for the initial and
      final mesons are $(\kappa_b,\kappa_c) = (0.089,0.119)$
      (solid circles) and $(0.119,0.089)$ (open circles).
      The light quark corresponds to the strange quark,
      $\kappa_l=0.1405$.
      Note that the lattice spacing~$a$ is held fixed here.
      }
    \label{fig:p2-extrapolation}
  \end{center}
\end{figure}

To obtain the value of
$h^{B\to D}_-(w)/h^{B\rightarrow D}_+(w)$ at the
zero-recoil limit, we extrapolate the plateau values of
$R^{(B\to D)/(D\rightarrow D)}_{V_i/V_0}$ for
${\vek{p}'}^2\to 0$.
The small piece of order $w-1$ vanishes in this limit as
well as the possible~$w$ dependence of form factors, so we obtain
$h^{B\to D}_-(1)/h^{B\rightarrow D}_+(1)$ without further
approximation.
Figure~\ref{fig:p2-extrapolation} shows the extrapolation for the same
mass values as in Fig.~\ref{fig:R_BDoverDD}.
There is no significant dependence on~$(a\vek{p}')^2$.
Thus, we simply apply a linear form to fit the data,
shown in the figure.
The numerical data in the zero-recoil limit are given in
Table~\ref{tab:h-_data}.
\begin{table}[tbp]
  \begin{center}
    \caption{Numerical data in the zero-recoil limit for
      $R^{(B\to D)/(D\rightarrow D)}_{V_i/V_0}$,
      which corresponds to $1-h_-(1)/h_+(1)$, at $\kappa_l=0.1405$.
      Rows (columns) are labeled by the value of $\kappa_h$ in the
      initial (final) state.
      Combinations without data have not been calculated in this work.
      The diagonal elements are~1 by construction.}
    \label{tab:h-_data}
\vskip0.3em
    \begin{tabular}{l|llllll}
\hline\hline
\multicolumn{1}{c|}{$\kappa_h$}
      & \multicolumn{1}{c}{0.062} & \multicolumn{1}{c}{0.089} &
        \multicolumn{1}{c}{0.100} & \multicolumn{1}{c}{0.110} &
        \multicolumn{1}{c}{0.119} & \multicolumn{1}{c}{0.125} \\
\hline
0.062 &1        &1.067(12)&1.093(14)&         &         &1.181(21)\\
0.089 &0.892(20)&1        &1.033(04)&1.063(08)&1.095(11)&1.121(15)\\
0.100 &0.836(27)&0.963(05)&1        &         &         &1.092(11)\\
0.110 &         &0.923(10)&         &1        &1.034(04)&         \\
0.119 &         &0.878(15)&         &0.964(04)&1        &         \\
0.125 &0.636(47)&0.837(20)&0.889(13)&         &         &1        \\
\hline\hline
    \end{tabular}
  \end{center}
\end{table}

\begin{figure}[btp]
  \begin{center}
    \epsfxsize=12cm \epsfbox{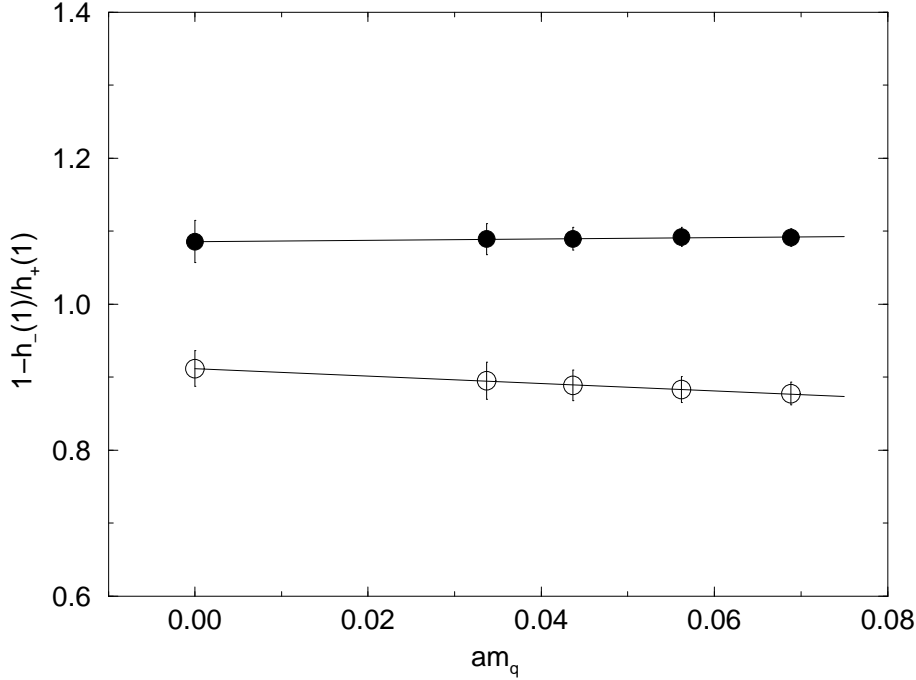}
    \caption{Chiral extrapolation of $1-h_-(1)/h_+(1)$.
      The heavy quark hopping parameters for the initial and
      final mesons are $(\kappa_b,\kappa_c) = (0.089,0.119)$
      (solid circles) and $(0.119,0.089)$ (open circles).}
    \label{fig:h-_chiral}
  \end{center}
\end{figure}

The chiral extrapolation of $1-h_-(1)/h_+(1)$ is
shown in Fig.~\ref{fig:h-_chiral}, for the combinations
$(\kappa_b,\kappa_c)=(0.089,0.119)$ and~$(0.119,0.089)$.
As in the case of $|h_+(1)|^2$, the dependence on~$am_q$ is
insignificant, but the (statistical) uncertainty increases,
by a factor of two.

\section{Heavy quark mass dependence of $h_-(1)$}
\label{sec:Heavy_quark_mass_dependence_of_h-}

As with~$h_+(1)$, the heavy quark mass dependence of~$h_-(1)$ can be
described, in the heavy quark limit of QCD, with a $1/m_Q$ expansion.
The form of the heavy quark expansion of~$h^{B\to D}_-(1)$ is
restricted by its anti-symmetry under the exchange of the
initial and final states,
\begin{equation}
  \label{eq:1/m-expansion_of_h-}
  h_-(1) = - \left( \frac{1}{m_c} - \frac{1}{m_b} \right)_3
  \left[ c_-^{(1)}
  - c_-^{(2)} \left( \frac{1}{m_c} + \frac{1}{m_b} \right)_2
  \right]
  + O(1/m_Q^3).
\end{equation}
The meaning of the subscripts on the combinations of inverse masses
is given below.
The ratio $h_-(1)/h_+(1)$ obeys the same expansion up to the given
order, since the correction to the $h_+(1)$ starts at order~$1/m_Q^2$.

To take radiative corrections into account, we should note that the 
(lattice) ratio~$R^{(B\to D)/(D\to D)}_{V_i/V_0}$ corresponds 
to~$[1-h_-/h_+]/\rho_{V_i}$.
The right-hand side of~(\ref{eq:1/m-expansion_of_h-}), on the other
hand, is justified in HQET when radiative corrections are ignored.
Thus, we should multiply the data of Table~\ref{tab:h-_data} by
$\rho_{V_i}/(1-\beta_V)$.
We shall not do this for two reasons.
First, the one-loop contribution to~$\rho_{V_i}$ is not yet available,
although a calculation is in progress~\cite{Kronfeld_Hashimoto_98}.
Second, there is an indication from an analysis of renormalons that the
series for $\beta_V$ converges poorly~\cite{Neubert_Sachrajda_94}.
With these points in mind, we omit radiative corrections and
employ~(\ref{eq:1/m-expansion_of_h-}) as an Ansatz for interpolation.

The subscripts on the parentheses in~(\ref{eq:1/m-expansion_of_h-})
mean that the enclosed masses should be taken to be~$m_3$ or~$m_2$,
introduced in Sec.~\ref{sec:HQET_and_1/m_Q_expansion}.
The reasoning is as follows.
The contribution to~$h_-(1)$ of first order in~$1/m_Q$ comes solely from
the current~\cite{Luke_90}, namely the $1/m_3$ terms in~(\ref{eq:Vhh}).
The second-order contribution comes mainly from the first-order
contribution iterated with the $1/m_Q$ corrections to the
Hamiltonian~\cite{Falk_Neubert_93,Kronfeld_99}, namely the $1/m_2$
and $1/m_B$ terms in~(\ref{eq:H}).
We can take $m_B=m_2$ because, with the clover action, the difference
affects the interpolation negligibly.
Tracing the $1/m_Q$ expansion in this way, and making use of the 
anti-symmetry under the exchange of initial and final states, leads to 
the heavy-quark expansion for the lattice data of the form given 
in~(\ref{eq:1/m-expansion_of_h-}).

\begin{figure}[btp]
  \begin{center}
    \epsfxsize=12cm \epsfbox{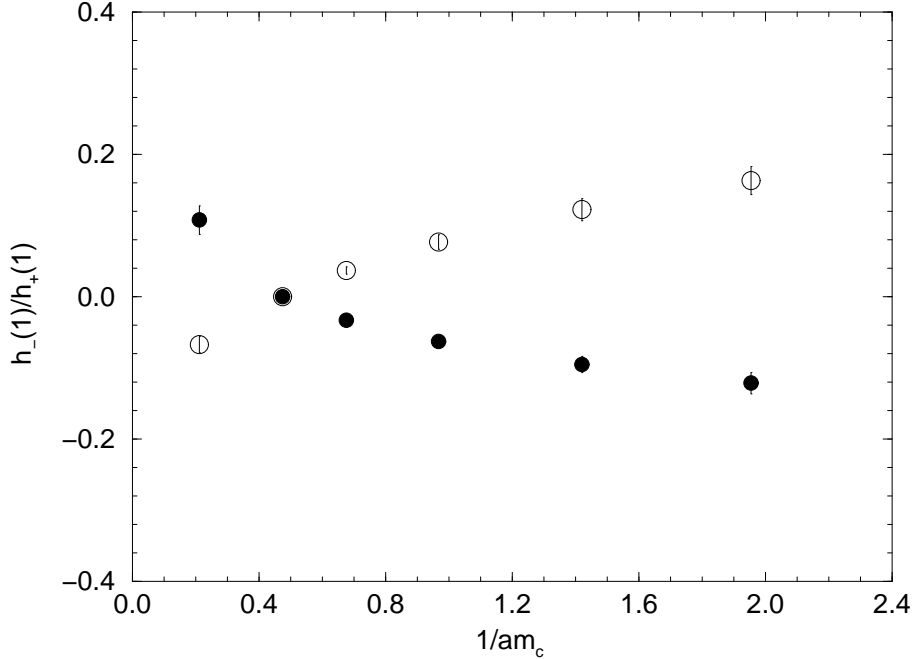}
    \caption{$1/am_c$ dependence of $h_-(1)/h_+(1)$.
      The initial heavy quark mass is fixed at
      $\kappa_b=0.089$ (solid circles), which corresponds
      to $1/am_b=0.475$.
      The open circles are obtained by exchanging the
      initial and final states.
      The light quark corresponds to the strange quark,
      $\kappa_l=0.1405$.}
    \label{fig:h-_heavy_mass_dependence}
  \end{center}
\end{figure}

In Fig.~\ref{fig:h-_heavy_mass_dependence} we plot the
$1/am_c$ dependence of $h_-(1)/h_+(1)$.
The solid circles are obtained by fixing the initial-state quark
mass to be $1/am_b=0.475$ and varying the final-state mass.
The open circles are obtained by fixing the final-state mass and
varying the initial-state mass.
We can clearly observe the mass dependence, which makes it
possible to extract the value of the form factor for physical
masses.

\begin{figure}[btp]
  \begin{center}
    \epsfxsize=12cm \epsfbox{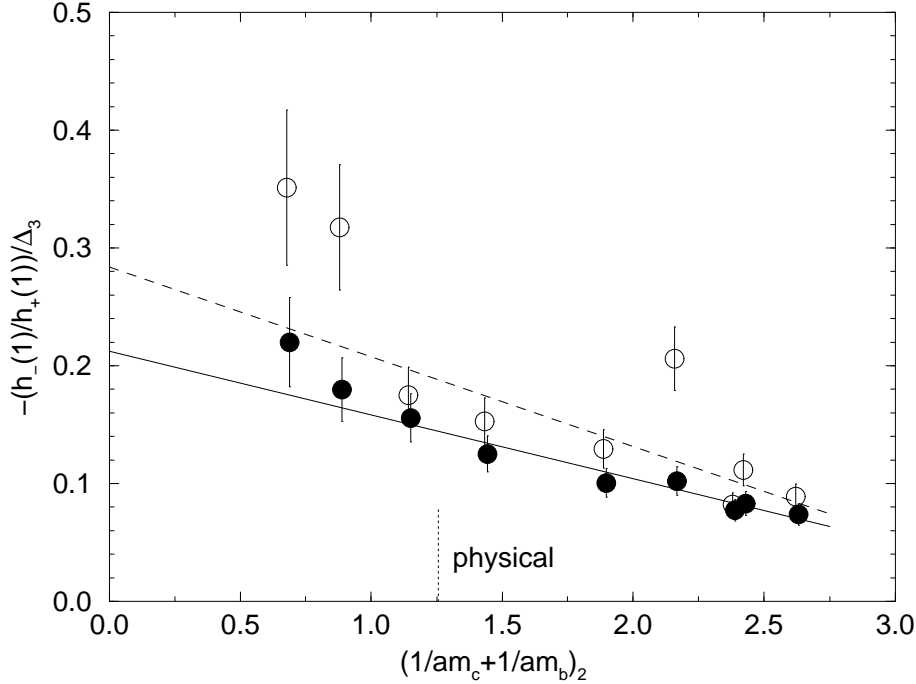}
    \caption{
      $-[h_-(1)/h_+(1)]/\Delta_3$ vs.\ $1/am_c+1/am_b$.
      Filled (open) symbols represent the ``heavier-to-lighter''
      (``lighter-to-heavier'') decay results.
      The solid and dashed lines are fitted results to the
      solid and open data points, respectively.
      The dotted vertical line indicates the physical value of
      $1/am_c+1/am_b$.
      The light quark corresponds to the strange quark,
      $\kappa_l=0.1405$.}
    \label{fig:h-_heavy_mass_coeff}
  \end{center}
\end{figure}

To extract the coefficients $c_-^{(1)}$ and $c_-^{(2)}$ we plot
in Fig.~\ref{fig:h-_heavy_mass_coeff}
\begin{equation}
  \label{eq:m-dependence_of_h-_reduced}
  \frac{R^{(B\to D)/(D\to D)}_{V_i/V_0}-1}{\Delta_3} =
  - \frac{h_-(1)/h_+(1)}{\Delta_3} =
  c_-^{(1)} - c_-^{(2)}
  \left( \frac{1}{am_c} + \frac{1}{am_b} \right)_2,
\end{equation}
where now $\Delta_3=1/am_{3c}-1/am_{3b}$.
Here the solid symbols represent the results from the
``heavier-to-lighter'' transitions and the open symbols from
``lighter-to-heavier'' transitions.
The two sets of data are consistent with each other, except three
points appearing well above the other points.
These data involve the heaviest quark mass in our calculation, where
the statistical noise is very large, and reliable fits become difficult.
The data are well described by the linear
form~(\ref{eq:m-dependence_of_h-_reduced}), and our results for its
coefficients extracted with the ``heavier-to-lighter'' data are
  $c_-^{(1)}=0.212(31)$ and $c_-^{(2)}=0.054(11)$.
In physical units, these coefficients are $c_-^{(1)}=0.246(37)$~GeV and
$c_-^{(2)}=[0.27(3)~\mathrm{GeV}]^2$.

The data presented in the figures and in Table~\ref{tab:h-_data} are
obtained with the rotation parameter~$d_1=0$.
One expects~$h_-$ to be sensitive to~$d_1$, because~$d_1$ is the
coefficient of an operator of order~$v$, and~$h_-$ parametrizes
a matrix element of order~$v$.
From the discussion in Sec.~\ref{sec:Lattice_and_HQ}, however, one 
sees that~$d_1$ influences matrix elements through the mass~$m_3$.
Thus, our method of fitting compensates for the omitted rotation,
provided we reconstitute the physical value of~$h_-(1)$ using the
physical values of the quark masses throughout.
A bonus of this method is that the radiative correction
factor~$\rho_{V_i}$ will be easier to compute when~$d_1=0$.

We have checked the influence of the rotation by repeating the
calculations with~$d_1=\tilde{d}_1$, cf.~(\ref{eq:tilde_d_1}).
The primary effect of varying~$d_1$ is through~$1/m_3$ and,
from~(\ref{eq:m3}) and~(\ref{eq:1/m-expansion_of_h-}),
is proportional to the difference $d_1^c-d_1^b$.
A secondary effect is to modify the radiative corrections of the lattice
currents.

With hopping parameters $(\kappa_b,\kappa_c)=(0.089,0.119)$, the
difference $\tilde{d}_1^c-\tilde{d}_1^b$ nearly vanishes.
Nevertheless, we find
\begin{equation}
    \label{eq:h_-_d_1}
    R^{(B\to D)/(D\to D)}_{V_i/V_0}(\tilde{d}_1) -
    R^{(B\to D)/(D\to D)}_{V_i/V_0}(0) = 0.0089\pm 0.0012,
\end{equation}
where we use the bootstrap method to obtain a statistical
uncertainty that takes correlations into account.
This difference must stem almost entirely from a change in the
radiative corrections, because the change in the heavy quark expansion
is, fortuitously, negligible.
Thus, it provides an estimate of the uncertainty from omitting the
radiative corrections.

%
%
%

Another check on the magnitude of the radiative corrections comes from
comparing the heavier-to-lighter transition with the lighter-to-heavier.
Because the physical form factor~$h_-$ is anti-symmetric under
interchange of the initial and final states, the incomplete
anti-symmetry of $R^{(B\to D)/(D\to D)}_{V_i/V_0}-1$, seen in
Table~\ref{tab:h-_data}, can come only from radiative corrections.
Near the physical region, these discrepancies are 10--20~\% of~$h_-(1)$.
With these considerations to guide an estimate, we take the uncertainty
in~$h_-(1)$ owing to unknown radiative corrections to range
from~$+0.010$ to~$-0.030$.

\section{Comparison with the QCD sum rules}
\label{sec:Comparison_with_the_QCD_sum_rules}

In the past, the form factors $h_+(1)$ and $h_-(1)$ have been
studied with QCD sum rules or the non-relativistic quark model.
Here we make a comparison of our results for $c_+^{(2)}$
and~$c_-^{(1)}$ with estimates obtained with those techniques.

From the zero-recoil sum rule, Shifman \textit{et al.}
obtain~\cite{Shifman_Uraltsev_Vainshtein_95}
\begin{equation}
  \label{eq:zero_recoil_sum_rule}
  F_{B\to D}^2 + \sum_{X} F_{X}^2
  = 1 - \frac{\mu_{\pi}^2 - \mu_G^2}{4}
    \left( \frac{1}{m_c} - \frac{1}{m_b} \right)^2,
\end{equation}
where $F_{B\to D}$ corresponds to $h_+(1)$ and
the $F_{X}$ represent contributions of higher excited
states.
The hadronic parameters $\mu_{\pi}^2$ and $\mu_G^2$ are
estimated with other sum rules, and recent results are
$\mu_{\pi}^2=0.5(1)~\mathrm{GeV}^2$ and
$\mu_G^2=0.36~\mathrm{GeV}^2$~\cite{Bigi_Shifman_Uraltsev_97}.
The relation~(\ref{eq:zero_recoil_sum_rule}) gives an upper
bound for $h_+(1)$,
\begin{equation}
  \label{eq:sum_rule_bound}
  h_+(1) < 1- \frac{\mu_{\pi}^2 - \mu_G^2}{2}
    \left( \frac{1}{m_c} - \frac{1}{m_b} \right)^2,
\end{equation}
provided that the contributions~$F_{X}^2$ of higher excited states
are strictly positive.
This can be translated as a lower bound for the
coefficient~$c_+^{(2)}$:
\begin{equation}
  \label{eq:sum_rule_bound_for_the_coefficient}
  c_+^{(2)} > \frac{1}{2}(\mu_{\pi}^2-\mu_G^2)
            = (0.26^{+0.09}_{-0.12} \mathrm{GeV})^2.
\end{equation}
Our result $c_+^{(2)}=[0.20(4)~\mathrm{GeV}]^2$ is lower
than the central value but still consistent within errors.

In~\cite{Falk_Neubert_93,Neubert_94} the authors used the
non-relativistic quark model to estimate the coefficient~$c_+^{(2)}$.
Their results scatter in a range
(0.2--$0.4~\mathrm{GeV}^2)$, strongly depending on the
assumed shape of the quark-antiquark wave function and the
value of the valence light quark mass.

The form factor $h_-(1)$ has been studied with QCD
sum rules~\cite{Neubert_92b,Ligeti_Nir_Neubert_94}.
Applying their analysis to the heavy quark
expansion~(\ref{eq:1/m-expansion_of_h-}) one finds
\begin{equation}
  \label{eq:h-_heavy_quark_expansion}
  c_-^{(1)} = \frac{\bar{\Lambda}}{2}
             \left[1+\delta_1 - 2 (1+\delta_2) \eta(1)\right],
\end{equation}
where $\bar{\Lambda}=m_B-m_b$, the~$\delta_i$ are radiative
corrections, and $\eta(1)$ represents a ratio of HQET form factors,
at zero recoil.
Neglecting radiative corrections, Neubert~\cite{Neubert_92b} finds
$\eta(1)=1/3$ from a QCD sum rule.
Taking $\bar{\Lambda}=0.5\pm 0.1$~GeV and $\delta_1=\delta_2=0$,
this implies $c_-^{(1)}=0.08(2)~\textrm{GeV}$.
With radiative corrections in the sum rule, Ligeti \textit{et al.}\ 
find $\eta(1)=0.6\pm 0.2$~\cite{Ligeti_Nir_Neubert_94}.
Taking now $\delta_1=0.11$ and $\delta_2=0.09$~\cite{Neubert_93},
this implies $c_-^{(1)}=-0.05(10)~\mathrm{GeV}$.
Our result is significantly larger than both, but it is difficult to
make a direct comparison.
Our lattice calculation contains some of the radiative corrections
automatically, and the remainder has not yet been calculated.
When the lattice one-loop calculation is available, it should be
possible to make a direct comparison.
As we mentioned above, it is conceivable that these effects could change
$c_-^{(1)}$ significantly, without a great effect on the value we
extract for~$h_-(1)$.

\section{Result for $\mathcal{F}_{B\to D}(1)$}
\label{sec:Result_for_F}

In the previous sections we have investigated the heavy quark
mass dependence of $h_+(1)$ and $h_-(1)$ and obtained the
coefficients in the $1/m_Q$ expansions~(\ref{eq:1/m-expansion_of_h+})
and~(\ref{eq:1/m-expansion_of_h-}).
To extract the value of $\mathcal{F}_{B\to D}(1)$ we input the
physical values of $m_c$ and $m_b$, which we adjust to give the
physical meson masses.
At $\beta=5.7$ these parameters are $am_c=1.0(1)$ and $am_b=3.9(5)$.
The central value is fixed with the $D$ and $B$ meson masses with the
lattice spacing $a^{-1}(\mbox{1S--1P})$, and the error range reflects
the uncertainty in the lattice spacing.

The values of physical $h_+(1)$ and $h_-(1)$ (without the matching
factors) are given in Table~\ref{tab:form_factors} for three possible
combinations of $am_b$ and~$am_c$.
\begin{table}[tbp]
\begin{center}
\caption{Tree level estimate of the form factors at zero recoil,
with statistical errors only.
The light quark corresponds to the strange quark,
$\kappa_l=0.1405$.
The entries for the form factors do not reflect radiative corrections.}
\label{tab:form_factors}
\vskip0.3em
\begin{tabular}{ccccc}
\hline \hline
$am_b$ & $am_c$ & $h_+(1)$ & $h_-(1)$     & $\mathcal{F}_{B\to D}(1)$\\
\hline
4.4    & 1.1    & 0.992(3) & $-$0.103(13) & 1.041(8) \\
3.9    & 1.0    & 0.991(3) & $-$0.107(14) & 1.042(8) \\
3.4    & 0.9    & 0.990(4) & $-$0.112(14) & 1.043(8) \\
\hline \hline
\end{tabular}
\end{center}
\end{table}
Since the systematic errors in $am_b$ and in $am_c$ are
correlated, we consider the central and two limiting
combinations only.
The statistical errors on~$h_+(1)$ and~$h_-(1)$ are estimated with
the jackknife method, so that the resulting precision is better than
that obtained by adding in quadrature the errors on
coefficients~$c_{\pm}^{(n)}$.
In the physical amplitude $\mathcal{F}_{B\to D}(1)$, which is
the linear combination of $h_+(1)$ and $h_-(1)$ given
in~(\ref{eq:form_factor_relation}), the uncertainty from adjusting
the quark masses largely cancels, and the value
of~$\mathcal{F}_{B\to D}(1)$ is very stable.

To obtain the physical result, we must now fold in the radiative 
correction~$\rho_{V_0}$, relating the lattice current to the continuum.
Two of us recently have calculated this factor to one
loop~\cite{Kronfeld_Hashimoto_98}, and at $am_b=3.9$ and $am_c=1.0$
they find $\rho_{V_0}=1+0.096\alpha_s$.
The Lepage-Mackenzie scale $q^*$ for the coupling
$\alpha_s(q^*)$~\cite{Lepage_Mackenzie_93} has also been calculated,
and at the same quark masses the result is $q^*=4.4/a$.
At $\beta=5.7$, $\alpha_V(4.4/a)=0.168$ and the
correction to $h_{+}(1)$ is $+0.016(3)$, taking the error
of omitting higher orders to be 20~\% of the one-loop
correction.

A similar one-loop calculation for ~$\rho_{V_i}$, which modifies 
$h_-(1)$, is not yet available.
We allow, therefore, a systematic uncertainty for this effect.

Our results for the form factors are
\begin{eqnarray}
  \label{eq:result_h+}
  h_{+}(1) & = & +1.007 \pm 0.006 \pm 0.002 \pm 0.003, \\
  \label{eq:result_h-}
  h_{-}(1) & = & -0.107 \pm 0.028 \pm 0.004^{+0.010}_{-0.030},
\end{eqnarray}
where the error estimates are as follows.
The first error comes from statistics, after the chiral extrapolation;
the second from adjusting the heavy quark masses; and the third error
from unknown radiative corrections, two loops and higher for~$h_+$ 
and one loop and higher for~$h_-$.
The chiral extrapolations, which are shown in Figs.~\ref{fig:h+_chiral}
and~\ref{fig:h-_chiral}, double the statistical errors of
Table~\ref{tab:form_factors}, without changing the central values.

Our main result is the value of the form factor entering
the decay rate, at zero recoil.
Inserting the physical values of the $B$ and $D$ meson masses
and the results~(\ref{eq:result_h+}) and~(\ref{eq:result_h-})
into~(\ref{eq:form_factor_relation}),
\begin{equation}
  \label{eq:result_F(1)}
  \mathcal{F}_{B\to D}(1) = 
  	1.058 \pm 0.016 \pm 0.003^{+0.014}_{-0.005},
\end{equation}
where errors are from statistics, heavy quark masses, and omitted 
radiative corrections.
The last of these could be reduced substantially by calculating the
radiative correction factor~$\rho_{V_i}$ to one loop.

Two sources of uncertainty have yet to be investigated carefully.
They are the dependence on the lattice spacing and the effects of the
quenched approximation.
From our experience with $f_B$~\cite{Fermilab_97,JLQCD_97}, we might
suppose that these effects are a few percent and $\sim15~\%$,
respectively.
The ratios have been constructed so that all sources of error,
including these, vanish for equal heavy quark masses.
It is, therefore, our expectation that these percentages
apply not to $\mathcal{F}(1)$ but to $\mathcal{F}(1)-1$.
That means that these two sources of error should be under good
control, just as we have found with the other sources of uncertainty.

\section{Conclusions}
\label{sec:Conclusions}

In this paper we have shown that precise lattice calculations of
the zero-recoil form factors $h_+(1)$ and $h_-(1)$ are possible.
The principal technical advance is to consider ratios of matrix
elements, in which a large cancellation of statistical and systematic
errors takes place.
The numerical data are interpreted in a way mindful of heavy quark
symmetry~\cite{El-Khadra_Kronfeld_Mackenzie_97}.
We find, therefore, that the dependence of the form factors on the
heavy quark mass is well described by $1/m_Q$ expansions, and we
obtain the coefficients in the expansions.

Our control over the heavy quark mass dependence allows us to determine
the individual form factors~$h_+(1)$ and~$h_-(1)$, as well as the
physical combination $\mathcal{F}_{B\to D}(1)$.
The main results~(\ref{eq:result_h+})--(\ref{eq:result_F(1)})
account for most uncertainties, but not the dependence on the lattice
spacing or the effect of the quenched approximation.
Since our method is designed to yield the deviation of
$\mathcal{F}_{B\to D}(1)$ from one, we do not expect these
qualitatively to spoil the quoted precision.
With the proof of principle provided by this work, it should be
possible, in the short term, to obtain $\mathcal{F}_{B\to D}(1)$
with control over all sources of uncertainty and an error bar that
is small enough to be relevant to the determination of~$|V_{cb}|$.

\section*{Acknowledgments}
We thank Zoltan Ligeti and Ulrich Nierste for comments on the
manuscript.
High-performance computing was carried out on ACPMAPS; we thank past
and present members of Fermilab's Computing Division for designing,
building, operating, and maintaining this supercomputer, thus making
this work possible.
Fermilab is operated by Universities Research Association Inc.,
under contract with the U.S.\ Department of Energy.
SH is supported in part by the Grants-in-Aid of the
Japanese Ministry of Education under contract No.~11740162.
AXK is supported in part by the DOE OJI program under contract
DE-FG02-91ER40677 and through the Alfred P. Sloan Foundation.

\end{document}